\documentclass[article]{JHEP3}

\usepackage{amsmath,epsfig}

\usepackage{amssymb,amsfonts}

\title{Aspects of Dirichlet S-branes}

\preprint{\hepth{0507059}\\LPTHE-05-16\\LPTENS-05-22}

\author{B.~Durin$^\clubsuit$,
B.~Pioline$^{\clubsuit\spadesuit}$
\\\\
$\clubsuit$~LPTHE, Universit\'es Paris 6 et 7, 4 place Jussieu, \\
75252 Paris cedex 05, France
\\ \\
$\spadesuit$~LPTENS, D\'epartement de Physique de l'ENS, 24 rue Lhomond,\\
75231 Paris cedex 05, France
\\ \\
{\tt E-mail:\\
bdurin@lpthe.jussieu.fr, pioline@lpthe.jussieu.fr}
}

\abstract{In this note, we discuss some features of the 
Dirichlet S-brane, defined as a Dirichlet boundary condition 
on a time-like embedding coordinate of open strings. We analyze
the Euclidean theory on the S-brane world-volume, and 
trace its instability to the infinite fine-tuning of the initial conditions
required to produce an infinitely extended space-like defect.
Using their equivalence under T-duality with D-branes with supercritical 
electric field, we argue that under generic perturbation,
S-branes turn into D-brane / anti-D-branes. We extract the imaginary part 
of the cylinder amplitude, and interpret its inverse as a ``decay length'', 
beyond which a pair of S-branes annihilates. Finally, we reconsider 
the boundary state of the Dirichlet S-brane and find that it is either
a solution of type II string theory with imaginary R-R fields, or 
a solution of type II$^*$ with real fields. This leaves the non-BPS S-branes 
as potentially physical solutions of type II string theory.}

\makeatletter
\renewcommand{\subsubsection}{\@startsection{subsubsection}{3}{0mm}{-\baselineskip}{0.5\baselineskip}{\normalfont\normalsize\it}}
\makeatother

\newcommand{\pa}{\partial}

\def\bea{\begin{eqnarray}}
\def\eea{\end{eqnarray}}
\def\be{\begin{equation}}
\def\ee{\end{equation}}
\def\ba{\begin{align}}
\def\ea{\end{align}}
\def\bse{\begin{subequations}}
\def\ese{\end{subequations}}
\def\bi{\begin{itemize}}
\def\ei{\end{itemize}}

\def\1F1{{}_1\!F_1}
\def\2F0{{}_2\!F_0}

\DeclareMathOperator{\im}{Im}

\begin{document}
\maketitle 


S-branes are generically defined as defects localized on a
space-like hypersurface. In string theory, this designation covers
a variety of different constructions: the Dirichlet S-brane,
defined as a space-like locus where open strings can end, is
the direct counterpart of the D-brane \cite{Gutperle:2002ai}. The 
``decaying D-brane'' describes the rolling of the open string tachyon 
on an unstable D-brane to or from the closed string vacuum (see 
e.g. \cite{Sen:2004nf} for a review),
and in certain cases may be 
viewed as an array of Dirichlet S-branes in imaginary time 
\cite{Gaiotto:2003rm}. More generally, S-branes can be viewed 
as homogeneous time-dependent 
solutions in string theory or supergravity, primarily localized at
a given time (see e.g. \cite{Ohta:2004wk} for a review). 
Although they are expected to require very
fine-tuned initial conditions, they may be useful tractable models
of time-dependent backgrounds in string theory. Furthermore, 
S-branes may be important ingredients in constructing 
inflationary backgrounds in string theory  \cite{Hull:1998vg}, 
just as D-branes are useful sources of AdS spaces.
Finally, they may be used as probes \cite{Nekrasov:2002kf} or 
initial/final value surfaces \cite{Nitti:2005ym} 
in more general time-dependent
backgrounds.

In this note, we study some aspects of Dirichlet S-branes, 
obtained by imposing a Dirichlet condition on the time coordinate
(or a time-like combination) of open strings. We re-analyze 
the world-volume theory of a single S-brane, and relate
the ``instability'' due to the ``wrong'' signature of the transverse
fluctuations to the infinite fine-tuning which is necessary in order to 
maintain an infinitely extended, flat S-brane under radial evolution.
We show that under T-duality with respect to a transverse
spatial coordinate, S-branes are T-dual to D-branes carrying a
supercritical electric field. By analogy with the discharge of the
supercritical electric field due to the nucleation and 
stretching of open strings,
this suggests that generic S-brane configurations may decay into
a shower of D-brane - anti-D-brane pairs. We study the first 
quantization of the open strings stretched between
two S-branes, and find that, in contrast to the D-brane case, there
are only a finite number of physical states, which grows as the
S-branes are further separated in time. We analyze the cylinder 
amplitude for two parallel S-branes, and interpret the infrared
divergences due to these physical states as another signal of 
this instability.

Although the flat S-brane configurations are infinitely fine-tuned,
it is important to note that their
mere existence may have a potentially disastrous
effect: since S-branes can be viewed as the world-volume of a tachyonic
particle, they may signal an instability of the theory at
hand. Indeed, S-brane may sometimes be viewed as the thin-shell
approximation of a tunneling event from a false to a true vacuum,
although this requires a non-standard analytic continuation from
the Coleman-de Luccia instanton \cite{Astefanesei:2005eq}. It is thus 
important to ascertain whether the S-brane configurations 
mentioned above are indeed
solutions of string theory, with the correct reality conditions
on the fields. As we shall explain, the boundary state for the 
Dirichlet S-brane constructed in \cite{Gutperle:2002ai}  is 
a solution of type II* \cite{Hull:1998vg} rather than
type II string theory\footnote{This issue was also raised by
the authors of \cite{Skenderis:2003da}, in a different, plane wave 
background.}. A S-brane type solution of type II string theory can be obtained
by analytical continuation from the D-brane boundary state but it radiates
Ramond-Ramond fields with the wrong reality property.
As for the ``decaying D-brane'' mentioned above, 
although it is possible in the bosonic string case to arrange the
deformation parameter so as to describe a Dirichlet S-brane at
$X^0=0$ (together with its translates in imaginary time), this is
not longer possible in the type II superstring, as it would require
an imaginary tachyon field \cite{Okuda:2002yd}. This 
leaves the possibility of ``non-BPS''
Dirichlet S-branes, which do not source any Ramond field.
We shall not study these ``non-BPS'' S-branes in this note.

\subsubsection*{S8-brane effective action : a hint for S-brane decay}

Let us start by analyzing the world-volume effective action 
for a single Dirichlet S8-brane. The latter can be obtained by
reducing the Born-Infeld action for a space-filling D9-brane
\be
\label{D9}
S_{D9} = \frac{1}{g_s} \int d^{10}x ~ \sqrt{-\det \eta_{\mu\nu} + F_{\mu\nu}}
\sim \frac{1}{g_s} \int d^{10}x ~\frac12 (\pa_0 A_i - \pa_i A_0)^2 - \frac14 
(\pa_i A_j - \pa_j A_i)^2 
\ee
to field configurations which are independent of $x^0$, 
as open strings carry no momentum along this direction. 
Renaming the component $A_0$ of the gauge field as a scalar field
$X^0$, we obtain a nine-dimensional Euclidean theory
\be
\label{S8}
S_{S8} \sim \frac{1}{g_s} \int d^9x ~\left(\frac12 (\pa_i X^0)^2 
- \frac14 F_{ij}^2\right) \ .
\ee
where $X^0(x^i)$ describes the fluctuations in the transverse 
time-like direction
$x^0$, and $F_{ij}$ is a magnetic field supported by the S-brane.
This action is unbounded both from below and from above, due to the
opposite effects of the gradient energy $(\pa_i X^0)^2$ 
and the magnetic energy. This indicates that the S8-brane 
worldvolume $X^0=cste$ is ``unstable'' towards the 
creation of large spatial gradients of $X^0$. This should however
be interpreted with some care, since this field theory exists only at 
a given time. Rather, one should think of it as a statistical field
theory at temperature $g_s$. The fact that the action is unbounded
both from below and from above means that the thermodynamical ensemble
is unstable, and that the partition function diverges, due to the
contribution of configurations with large gradients. Equivalently, we
may treat this  statistical field theory as a quantum field theory,
by quantizing along a spatial direction, say $x^1$. The above
instability will then manifest itself by the growth of spatial gradients
under evolution along $x^1$.

Let us now restore the Born-Infeld corrections to the S8-brane 
action. Setting the magnetic field $F_{ij}$ to zero for convenience,
we find
\be
\label{S8BI}
S_{S8} = - \frac{1}{g_s} \int d^9x \sqrt{ 1 - (\pa_i X^0)^2 } \ ,
\ee
a higher-dimensional generalization of the action of 
a relativistic tachyonic particle, with imaginary mass $m=i/g_s$. 
Simple solutions with SO($n$) symmetry can be obtained by solving 
\be
\frac{\pa_r X^0}{\sqrt{1-(\pa_r X^0)^2}} = \frac{a}{r^{n-1}}\ .
\ee
Up to a redefinition $X^0 \to -X^0$, $a$ can be chosen to be
positive. 
The solutions read
\begin{align}
X^0(r) &= r\, {}_2 F_1 \left(\frac12, \frac1{2n-2};\frac{2n-1}{2n-2};-\frac{r^{2n-2}}{a^2}\right) && n>1\\
X^0(r) &= \frac{a}{\sqrt{1+a^2}} r && n=1 \ .
\end{align}
and have been plotted on Figure \ref{solcusp}. 
In the co-dimension 1 case ($n=1$), 
$\pa_r X^0$ is simply constant on disjoint intervals, with arbitrary
slope $|\pa_r X^0|<1$. For $n>1$, 
the solutions reach constant $X^0$ at infinity and 
have a cusp  $|X^0|\sim r$ at $r=0$, where the space-like world-volume becomes 
null. In more general situations without symmetry, 
we may expect a disordered manifold with null defects. 
At this point, as it will become clear in the T-dual
picture discussed below, open strings become massless,
and the Born-Infeld action cannot be trusted any longer,
A natural conjecture is that the
S-brane world-volume becomes time-like and describes a
shower of D-brane and anti-D-branes. This is consistent with 
the analysis in \cite{Hashimoto:2004yi}, where solutions of the 
S-brane action \eqref{S8BI} including the magnetic field were interpreted
as the formation or annihilation of D-branes or strings.

\FIGURE{
\centerline{\epsfig{file=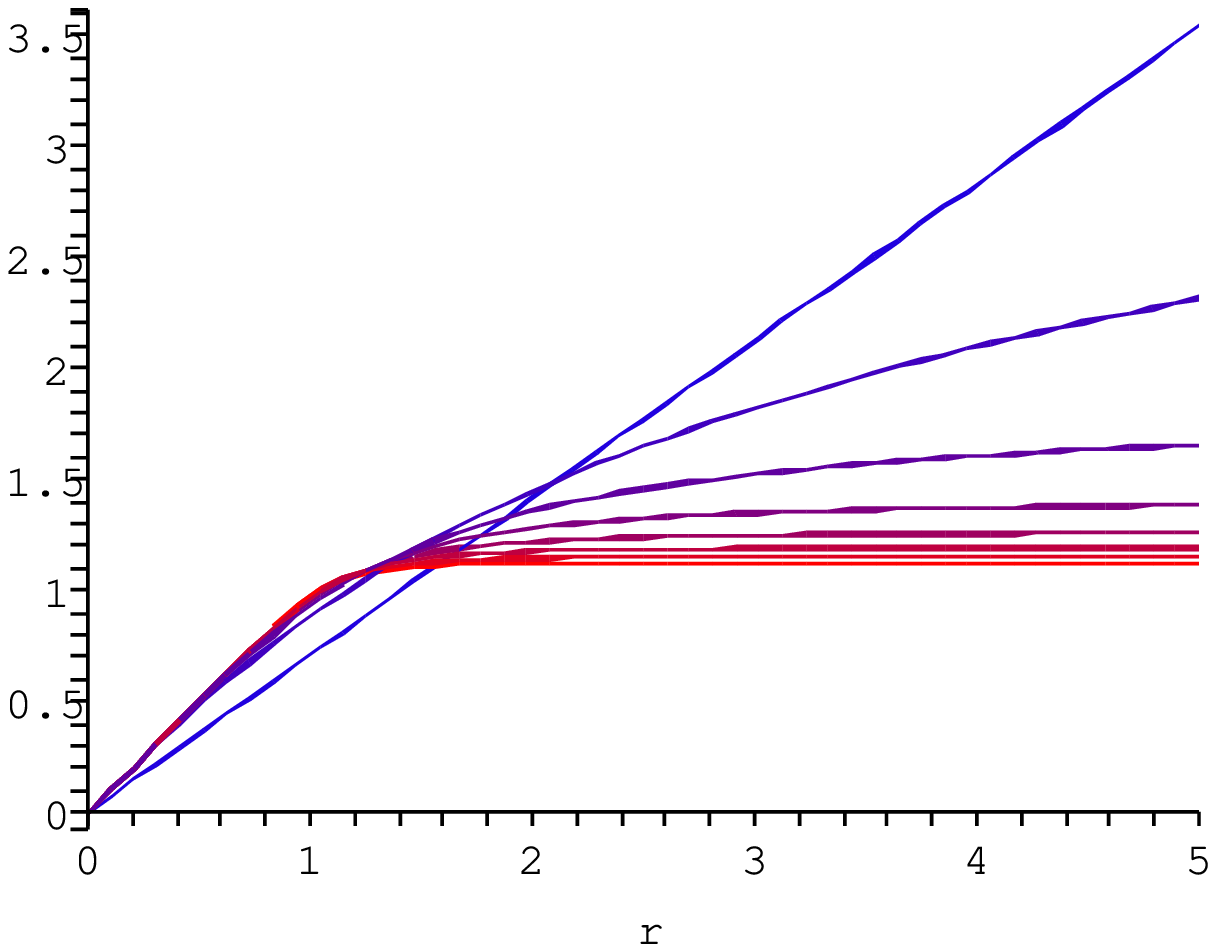, width=8cm}}
\caption{\label{solcusp} Profile of the transverse scalar $X^0$
on a S8-brane world-volume, as a function of the radial distance $r$
in  SO($n$) symmetric solutions ($a$ has been set to 1). The vertical axis is $X^0$. $n$ increases from $n=1$ (blue, top right) to  $n=8$ (red, bottom right).}
}

\subsubsection*{Space-like T-duality : from S-branes to open strings in a supercritical electric field}

In order to shed some light on the fate of S-branes, let us now study their
image under T-duality along a space-like coordinate. 
The boundary condition at $\sigma=0$ for a Dirichlet brane
moving along the direction $X^1$ at a velocity $v$
\be
\partial_\tau (X^1 -v X^0) =0 \ ,\quad  \partial_\sigma (X^0 -v X^1) =0 
\ee
For $|v|<1$, this corresponds to an ordinary D-brane, while for 
 $|v|>1$, this is an S-brane with tilted world-volume. 
The limiting case $\lvert v\rvert =1$ corresponds to a null brane.
Static D-branes can be obtained by setting $\lvert v\rvert=0$,
while S-branes at a fixed time correspond to $\lvert v\rvert \to \infty$.
T-dualizing along the spatial coordinate $X^1$, we are thus led to
\be
\partial_\sigma X^1 -v \partial_\tau X^0 =0\ ,\quad  \partial_\sigma X^0 
- v \partial_\tau X^1=0
\ee
This is now the boundary condition for a D-brane filling the
coordinates $(X^0,X^1)$, with electric field $F_{01}=v$ \cite{Bachas:1992bh}.
S-branes are thus T-dual to a D-brane with a {\it supercritical} field
$|F_{01}|>1$\footnote{In contrast, T-duality with respect to $X^0$
would lead to a sub-critical electric field $F_{01}=1/v$, but in
type II* rather than in type II string theory \cite{Hull:1998vg}.
The status of the former remains ill-understood.}. 
In such a supercritical field, the tension of open strings is insufficient
to overcome the pull from the electric field, and open strings
are nucleated from the vacuum and 
stretched to infinite length in finite time. In the limit $v \to \infty$
the semi-classical
string configuration describing this process is a strip of
finite extent along $X^0$ 
\be
X^0 = w \sigma\ ,\quad
X^1 = w \tau /F\ ,\quad
X^k = 2 \alpha' p_k \tau
\ee
where the mass shell condition requires
\be
w^2 = 4 \alpha' \left(\alpha' p_k^2 + (N-a)\right) F^2/(F^2-1) \ ,
\ee
where $N$ is the total excitation level and $a$ is the intercept
($a=1$ in the bosonic string, and $a=1/2$ or $a=0$ for the
superstring in the Neveu-Schwarz or Ramond sector). In particular,
the worldsheet coordinate $\sigma$ becomes identified with the 
target-space time $X^0$.
Allowing fluctuations away from this rigid configuration, 
this can be viewed as a set of dipoles nucleating from the
vacuum, stretching and rapidly merging into a single string of infinite length.

In general, we expect that such a creation will discharge the
capacitor plates and relax the electric field until 
it becomes critical or sub-critical. This suggests that
in the T-dual version, the creation of stretched strings under
evolution along a spatial direction $x^1$ will bend the S-branes
until they become null, and possibly turn into pairs of D-branes -
anti-D-branes.

\subsubsection*{Open strings stretched between two S-branes}

Let us now consider a configuration of 
two parallel Dirichlet S$p$-branes (at a fixed time).
An open string stretching between the two S-branes 
has the following mode expansion
\bea
\label{modxmu}
X^\mu(\tau, 0) &= x^\mu_0 + w^\mu \sigma+i\sqrt{2\alpha'}\sum_{n\neq0}\frac{a^\mu_n}{n} e^{-i n \tau} \sin (n\sigma) \\
\label{modxk}
X^k(\tau, 0) &= x^k_0 + 2\alpha' p^k \tau+i\sqrt{2\alpha'}\sum_{n\neq0}\frac{a^k_n}{n} e^{-i n \tau} \cos (n\sigma) \ ,
\eea
where $X^\mu$ denotes the time-like coordinate $X^0$
and the $D-p-2$ transverse spatial coordinates
 $X^i$; $X^k$ are the $p+1$ coordinates of the world-volume of the S-brane and
$w^\mu = (x^\mu_1-x^\mu_0)/\pi$
if we set the following boundary conditions
\be
X^\mu(\tau, 0) = x^\mu_0\  ,\   X^\mu(\tau,\pi) = x^\mu_1 \ .
\ee
In the rigid limit ($a^{\mu,k}_n =0$), much as in the super-critical electric
field case, this configuration describes an infinitely long string
which exists for a finite time interval only (see Figure
\ref{wldsheet}). Allowing for fluctuations,
this can again be viewed as a set of open string dipoles nucleating
from the vacuum shortly before the S-brane, and being stretched to
infinite size. The mass-shell condition requires
\be
p_k^2 = \frac{\Delta^2}{4\pi^2{\alpha'}^2} + \frac{a - N}{\alpha'}\ .
\ee
where $\Delta$ is the time-like separation between the two S-branes,
\be 
\Delta^2 = (x^0_1 - x^0_0)^2 - (x^i_1-x^i_0)^2\ .
\ee
For a given time-like distance $\Delta^2>0$, there are therefore only
a finite number of physical states with real transverse momenta
$p_k$. In the limit of coinciding S-branes, the only physical states
are the ``massless'' modes of \eqref{S8}, together with the tachyon
mode in the bosonic string case. As we shall see, these physical modes
are at the origin of infrared divergences in the one-loop amplitude,
which signal the fine-tuning of the flat S-brane configuration.

\FIGURE{
\epsfig{file=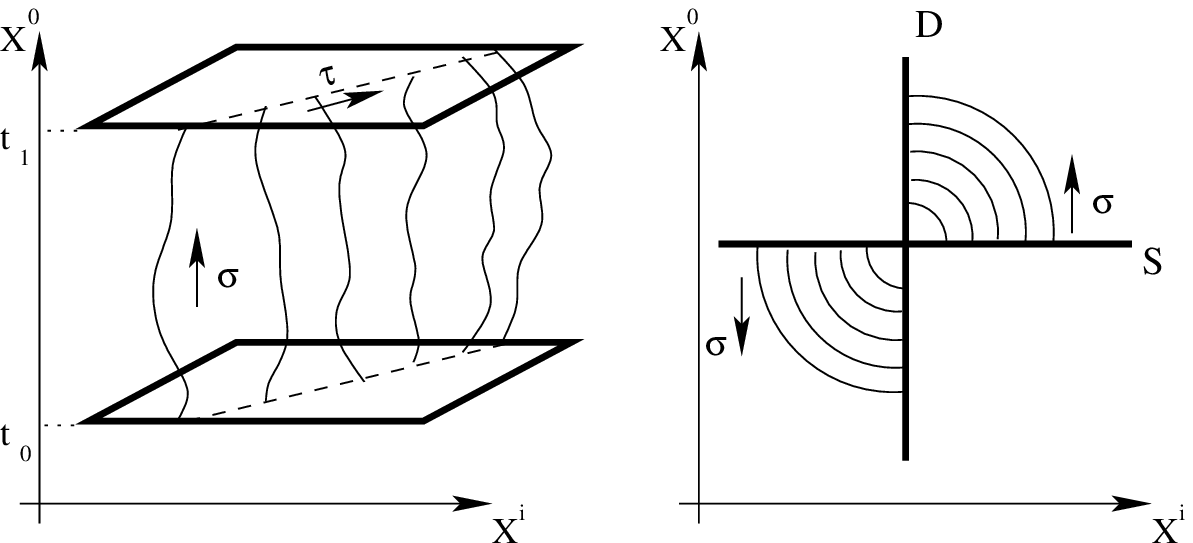, width=13.3cm}
\caption{\label{wldsheet} Classical worldsheet of an open string stretched between two S-branes (left) and one  S-brane and one D-brane (right). Dashed lines are the boundary of the worldsheet, solid lines stand for open strings at a given worldsheet time $\tau$, thick lines represent branes. For the S-D configuration, only the lowest open string mode is excited in our sketch.}
}

\subsubsection*{Open strings stretched between one S-brane and one D-brane}

Let us now focus on an open string stretched between a S8-brane 
at fixed time $x^0$ and a static D8-brane at position $x^1$ (see Figure \ref{wldsheet}). 
The boundary conditions 
\begin{align}
\partial_\tau X^0 &= 0  & \partial_\sigma X^1 &=0 && \text{S-brane at }\sigma = 0 \notag\\
\partial_\sigma X^0 &= 0  & \partial_\tau X^1 &=0 && \text{D-brane at }\sigma = \pi \ .
\end{align}
lead to the following mode expansion
\begin{align}
X^0 (\tau, \sigma) &= x^0 +\sqrt{2\alpha'} \sum_{n=-\infty}^\infty \frac{a_n^0}{n+1/2} e^{-i(n+1/2)\tau} \sin[(n+1/2)\sigma] \notag\\
X^1 (\tau, \sigma) &= x^1 +i\sqrt{2\alpha'} \sum_{n=-\infty}^\infty \frac{a_n^1}{n+1/2} e^{-i(n+1/2)\tau} \cos[(n+1/2)\sigma]\ .
\label{sdmexp}
\end{align}
where $(a_n^0)^* = -a_{-n-1}^0, (a_n^1)^* = a_{-n-1}^1$.
The transverse coordinates $X^k$ have of course the standard mode expansion
\eqref{modxk}. Note that this mode expansion is similar to that of
an open string stretched between a D$p$-brane and a D$p'$-brane.
Truncating to the lowest mode $n=0$, the classical open string worldsheet
in the $(x^0,x^1)$ plane is now the quadrant of a disk 
of finite radius centered at the intersection
between the S-brane and the D-brane. It can be interpreted as
a string dipole nucleating from the vacuum shortly before the S-brane,
whose one end stays on the D-brane but whose other end is stretched to 
a finite distance. It is tempting to speculate that the condensation of
the open string tachyon may lead to a reconnection of the S and D-brane.




\subsubsection*{Open string one-loop amplitude and S-brane correlation length}

We now turn to the one-loop amplitude of open strings stretched
between two S-branes. Before proceeding to the actual computation, 
let us discussing its physical interpretation. 

In the case of two static D-branes, the cylinder amplitude 
is equal to the energy of interaction between the two D-branes, 
times an infinite factor corresponding to the infinite duration $T$
of the interaction. This is because, in the closed string
channel, the cylinder amplitude is just the tree-level amplitude of
closed strings in the presence of static sources, schematically,
\be
A_{cyl} = \frac{i}2 \int d^D x\, d^D y\, J_1(x) G_F(x-y) J_2(y) \label{int}
= T E_{12}
\ee
where $G_F$ is the Feynmann propagator and 
\be
J_1(x^0, x^i)=q_1 \delta(x^i-x^i_1)\ ,\quad J_2= q_2 \delta(x^i-x^i_2)
\ee
are the sources describing the two D-branes at rest, and $E_{12}$
is the interaction energy between the two sources.
The second equality in \eqref{int} follows
by quantizing along the time direction $x^0$, and provides the
simplest way of extracting the Coulomb potential in QED. 
A non-vanishing imaginary part of the cylinder amplitude 
would imply that the interaction energy $E_{12}$
of the combined D-branes has an imaginary
part, and therefore that the overlap between the in-state and
out-state is zero, signalling D-brane decay after a characteristic
time $1/\im E[J]$. This approach was used in \cite{Bardakci:2001ck,
Craps:2001jp} to compute the decay rate of unstable D-branes.

In the case of the cylinder amplitude between two S-branes, 
a similar reasoning identifies $A_{cyl}$ with
the tree-level amplitude of closed strings in the background of
classical sources 
\be
J_1=q_1 \delta(x^0-x^0_1)\delta(x^i-x^i_1)\ ,\quad
J_2=q_2 \delta(x^0-x^0_2)\delta(x^i-x^i_2)
\ee
Time translation is of
course broken, but not translations along a common spatial direction
of two S-branes, say $x^k$. Thus, the cylinder amplitude computes the
product of an infinite {\it interaction length} $L_k$ by the {\it interaction
momentum} along $x^k$, due to the emission and reabsorption 
of closed strings by the two sources:
\be
A_{cyl} = L_k P^k_{12}
\ee
As in the static case, an imaginary part signals a decay 
along the evolution $x_k$, after a characteristic distance $1/\im P_{12}^k$.

Let us now proceed to compute the one-loop amplitude of open strings stretched
between two infinitely extended, flat S-branes at fixed times. 
With the notations of the previous
subsection, the cylinder amplitude in the open string channel 
reads
\be
\label{olSr} A_{cyl}= \frac12 V_{p+1} \int_0^{\infty} \frac{dt}{t} \int
\frac{d^{p+1} p_k}{(2\pi)^{p+1}} e^{- 2\pi \alpha't [ p_k^2 - 
\Delta^2/(2\pi \alpha')^2] } Z_{osc} 
\ee 
where the partition function of the oscillators is identical to the 
D-brane case \cite{Polchinski:1995mt},
\be
\label{zosc} 
Z_{osc} = \frac{-16 \prod_{n=1}^{\infty} (1+q^{2n})^8 
+ q^{-1} \prod_{n=1}^{\infty} (1+q^{2n-1})^8
- q^{-1} \prod_{n=1}^{\infty} (1-q^{2n-1})^8}
{\prod_{n=1}^{\infty} (1-q^{2n})^8}
\ee 
and $q = e^{-\pi t}$. In \eqref{zosc}, the three terms correspond to 
(one half) the
partition function of the open string oscillators in the Ramond sector, 
in the Neveu-Schwarz sector and in the Neveu-Schwarz sector with an insertion
of $(-1)^F$, respectively. 
In the dual channel $\tilde t=1/t$, the first two terms (the
third, resp.) correspond to the exchange of closed strings 
in the NS-NS (R-R, resp.) sector. As usual, $Z_{osc}$ vanishes due to Jacobi's
abstruse identity. We shall focus on the NS-NS part of the amplitude,
since we will argue later that the R-R part is unphysical and needs
to be projected out. 
The only difference with the D-brane cylinder amplitude 
lies in the zero-mode part. In contrast with the D-brane
case, the integral over the momenta $p_k$ is convergent without the need
for analytic continuation. As usual, the remaining $t$ integral diverges
in the ultraviolet ($t\to 0$), corresponding to the exchange of massless
closed strings. For null separation 
($\Delta=0$), the $t$ integral also diverges in the infrared, due
to the open string tachyon pole, but this is cancelled by the R-R part
of the amplitude.  When the time-like separation $\Delta$ increases,
however, more and more open string states become tachyonic, and lead
to ever more severe infrared divergences. The proper regularization of these
infrared divergences has been explained in \cite{Marcus:1988vs}, and leads
to an imaginary part for the one-loop amplitude\footnote{We used again the abstruse identity to equate the NS-NS part of the amplitude with minus the R-R part, which is simpler to deal with.}
\be 
\label{imas}
\im A_{cyl}^{\text{NS-NS}} = V_{p+1}\, 
{\cal N}  \sum_{n=-1}^{n^*} a_{n} 
\left(\frac{\Delta^2}{2\pi^2\alpha'}-n\right)^{\frac{p+1}2} \ee
where
\be {\cal N} = \frac{1}{2^{\frac{5+3p}2}\, \pi^{\frac{p-1}2}
{\alpha'}^{\frac{1+p}2} \Gamma \left(\frac{3+p}{2}\right)}\ ,
\ee
where $n^*$ is the greatest integer lower that $\Delta^2/(2\pi^2\alpha')$
and $a_n$ are the Fourier coefficients of the modular form 
\be
\frac{\vartheta_4^4}{\eta^{12}} 
=q^{-1}\prod_{n=1}^{\infty}\frac{(1-q^{2n-1})^8}{(1-q^{2n})^8}
:=\sum_{n=-1}^{\infty} a_n q^n  \qquad (q=e^{-\pi t})
\ee
These coefficients are well approximated by the leading term in
the Rademacher formula (see \cite{Dijkgraaf:2000fq} and references therein),
\be 
a_n \sim \pi (-1)^{n+1}  n^{-5/2} I_5 \left(2\pi \sqrt{n}\right) \sim (-1)^{n+1} n^{-11/4} e^{2\pi\sqrt{n}}
\ee
In particular, they alternate in sign.
Due to the Hagedorn growth 
of $a_n$, the sum in \eqref{imas} is dominated by the last non-vanishing
term,
\be
\label{imae}
\lvert\im A_{cyl}^{\text{NS-NS}}\rvert \sim \Delta^{-\frac{11}2} 
e^{\Delta\sqrt{2/\alpha'}}\ .
\ee
When $\Delta^2/(2\pi^2\alpha')$ is integer-valued, the last term 
vanishes, and the sum is dominated by the penultimate term $n=n^*-1$,
whose sign depends on the parity of $n^*$. The imaginary part of the
amplitude thus oscillates in sign with an overall growth
determined by \eqref{imae} (see Figure \ref{imfig}). In particular, it vanishes infinitely
many times, which raises the intriguing possibility that infinitely
extended S-brane pairs may exist at highly fine-tuned time-like separations.
In general however, the spatial decay takes place extremely rapidly,
on a characteristic distance of string scale.

\FIGURE{
\epsfig{file=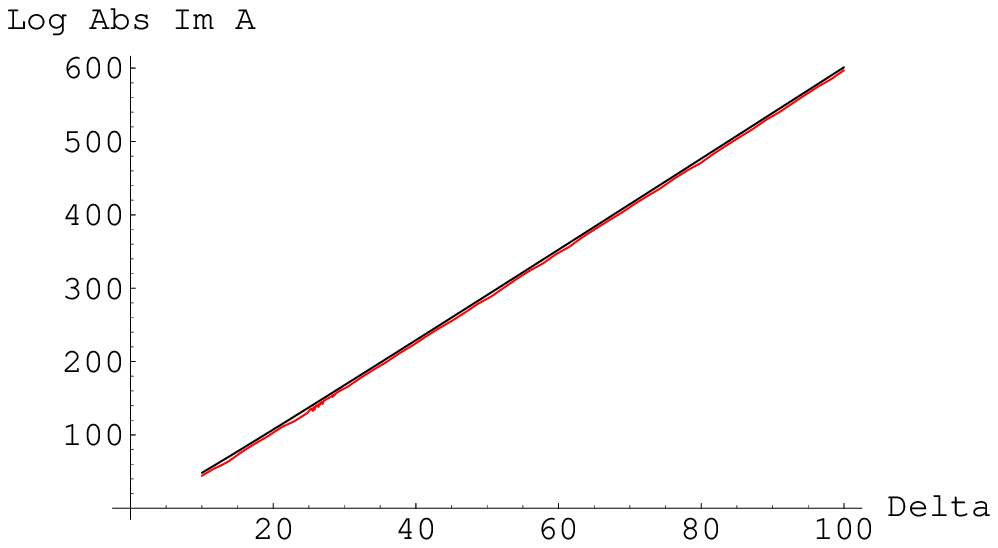, width=6.5cm} \ \ \ \ \ \ \ 
\epsfig{file=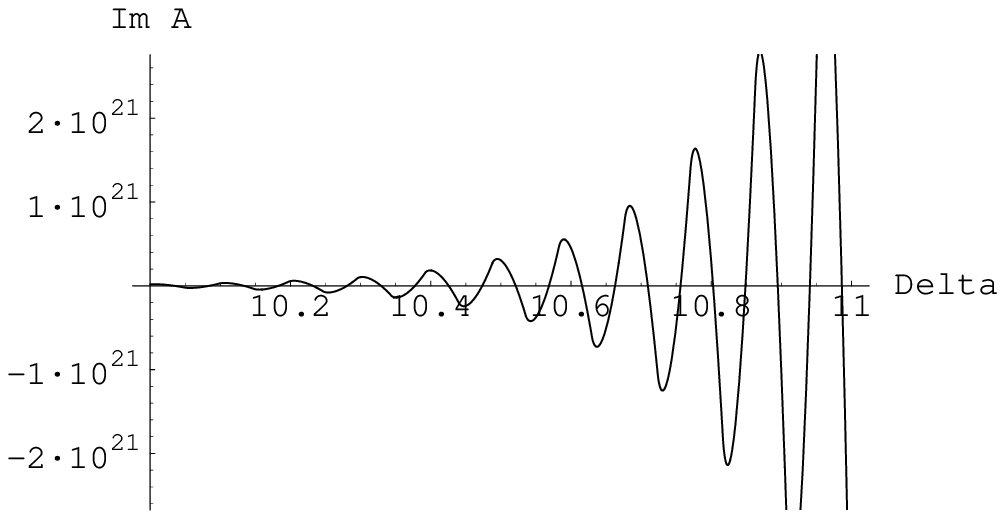, width=6.5cm}
\caption{\label{imfig} Left: $\log \lvert \im A_{cyl}^{\text{NS-NS}} \rvert$
 (in red) as function of $\Delta$. The curve cannot be distinguished from
its approximation \eqref{imae} (in black),
but misses the oscillations. Right: a zoom on the interval $\Delta\in[10,11]$
clearly shows the oscillations. We have set $p=2$ 
and $\alpha' = 1/(2\pi)$ for both plots.}
}

\subsubsection*{Boundary states for BPS S-branes}

As we have explained in the introduction, S-branes are the tachyonic
analogues of D-branes, hence their existence in type II string theory
would potentially be a disaster. We 
now return to the S-brane boundary state proposed in 
\cite{Gutperle:2002ai}, and investigate whether this boundary state 
is a solution of type II string theory, with the correct reality
conditions on the emitted closed string fields.
Following the conventions of \cite{DiVecchia:1999rh, 
Gaberdiel:2000jr}, the general D-brane boundary state is 
constructed from the Ishibashi states
\be 
\lvert B, \eta \rangle_{NS-NS} = {\cal N}_{NS-NS}\, \delta^{(d_\perp)} (q-y)\, e^{-\sum_{n=1}^\infty \alpha_{-n}^\mu S_{\mu\nu} \tilde \alpha_{-n}^\nu}\, e^{i \eta\sum_{r=1/2}^\infty \psi_{-r}^\mu S_{\mu\nu} \tilde \psi_{-r}^\nu }\, \lvert 0; k=0 \rangle
\ee
\be
\lvert B, \eta \rangle_{R-R} = {\cal N}_{R-R}\, \delta^{(d_\perp)} (q-y)\, e^{-\sum_{n=1}^\infty \alpha_{-n}^\mu S_{\mu\nu} \tilde \alpha_{-n}^\nu}\, e^{i \eta\sum_{n=1}^\infty \psi_{-n}^\mu S_{\mu\nu} \tilde \psi_{-n}^\nu}\, {\cal M}_{AB}  \lvert A\rangle \lvert \tilde B\rangle
\ee
where $(\alpha_n,\psi_r,\psi_n)$  are the left-moving modes
of the bosonic coordinates $X^\mu$ and fermionic coordinates 
$\psi^\mu$ in the NS and R sectors, $(\tilde\alpha_n,\tilde\psi_r,
\tilde\psi_n)$ are their right-moving partners, 
$q^a,y^a$ are the transverse position operator and eigenvalue,
$\lvert 0; k=0 \rangle$ is the NS-NS vacuum,  
$\lvert A\rangle \lvert \tilde B \rangle$ is the R-R vacuum
with $SO(1,9)$ spinor indices A and B. The matrix $S^\mu_\nu$
has eigenvalues $+1$ in the Neuman directions and $-1$ in the 
Dirichlet directions. The matrix ${\cal M}_{AB}$ is a solution
of the algebraic equation
\be\label{A10} (\Gamma^\mu)^{\text{T}} {\cal M} - i\eta S^\mu_{\phantom{\mu}\nu} \Gamma_{11} {\cal M} \Gamma^\nu = 0 
\ee
and can be chosen as 
\be
\label{mdd}
 {\cal M}_{\text{D-brane}} = C \Gamma^0 \Gamma^1
\ldots \Gamma^{p} \frac{1+ i\eta \Gamma_{11}}{1 + i\eta}\ .
\ee
GSO-invariant boundary states are constructed as superpositions
\be \lvert B \rangle_{NS-NS} = \lvert B, +\rangle_{NS-NS} 
- \lvert B, - \rangle_{NS-NS}\ee
and
\be \lvert B \rangle_{R-R} = \lvert B, + \rangle_{R-R} + \lvert B, - \rangle_{R-R}\ .\ee
The boundary state for a BPS D-brane is obtained by combining
NS-NS and R-R sectors 
\be 
\lvert D \rangle_{\text{BPS}} = \lvert B \rangle_{NS-NS} 
+ \lvert B \rangle_{R-R} 
\ee
In order to satisfy open-closed duality with a tachyon-free 
open-string spectrum, one should further require
\be ({\cal N}^D_{NS-NS})^2 = - \frac1{16}({\cal N}^D_{R-R})^2 \label{Dcd}
\ee
On the other hand, non-BPS branes only source NS-NS closed strings,
\be 
\lvert D \rangle_{\text{non-BPS}} = \lvert B \rangle_{NS-NS} 
\ee

The S-brane boundary state can be obtained from this general formalism
by setting
\be\label{S} S^\mu_{\phantom{\mu}\nu}=(-1,1,1,\ldots,1,-1,-1,\ldots,-1) 
\ee
and choosing a solution for \eqref{A10}. One option is to analytically
continue the D-brane solution \eqref{mdd}, leading to 
\be
\label{mds}
{\cal M}_{\text{S-brane}} = i C \Gamma^1 \Gamma^2
\ldots \Gamma^{p+1} \frac{1+ i\eta \Gamma_{11}}{1 + i\eta} 
\ee 
Due to the factor of $i$ in \eqref{mds}, if we keep the normalization 
condition \eqref{Dcd}, we obtain a S-brane with a GSO-odd projection
on the open string spectrum, in particular with an open string ``tachyon''.
We will denote this as a $S^-$-brane. It is easy to check from
the formulas 
in \cite{DiVecchia:1997pr} that, when $p<8$, 
the NS-NS closed string fields emitted 
by this boundary state are imaginary, while the R-R fields are real
\footnote{The
Fourier transform from momentum space to real space introduces a
factor of $i$, for $p<8$.}. In contrast, for the codimension 1 case ($p=8$), 
the NS-NS and R-R closed fields are real and imaginary, respectively.

\begin{table}
\begin{center}
\begin{tabular}{|c|c|c|c|c|c|}
\hline
 & D-brane & \multicolumn{2}{|c|}{S$^+p$-brane} & \multicolumn{2}{|c|}{S$^-p$-brane} \\
\cline{3-6} 
 & & $p<8$ & $p=8$ & $p<8$ & $p=8$ \\
\hline
NS-NS field & $\mathbb{R}$ & $i\mathbb{R}$ & $\mathbb{R}$ & $i\mathbb{R}$ & $\mathbb{R}$ \\
\hline
R-R field & $\mathbb{R}$ & $i\mathbb{R}$ & $\mathbb{R}$ & $\mathbb{R}$ & $i\mathbb{R}$ \\
\hline
\end{tabular}
\end{center}
\caption{Reality properties of the closed string fields radiated by 
D-branes and $S^\pm$-branes.}
\end{table}

On the other hand, since \eqref{A10} is invariant under rescalings 
of ${\cal M}$, we may choose to drop the factor of $i$ on the right-hand
side, or, equivalently,  flip the sign on the right-hand side of \eqref{Dcd}.
This corresponds to a S-brane with a GSO-even projection, which we will
refer to as a $S^+$-brane: this is the choice made in \cite{Gutperle:2002ai}.
Both the NS-NS and  R-R fields sourced by the $S^+$-brane are now 
purely imaginary (except for $p=8$,
where they are both real). These results are summarized on Table 1.
For $p<8$, purely real fields may be obtained
by multiplying the boundary state by an overall factor of $i$.

However, while analytic continuation from the D-brane case guarantees
that the $S^-$-brane is still a solution of the equations of motion
(albeit with unphysical fields), this is not the case for the $S^+$-brane.
Indeed, there are additional consistency conditions which follow from
the supergravity equations of motion. Schematically, the gravitational
and gauge fields radiated by a boundary state $|S\rangle$
can be expanded in powers
of the string coupling as
\begin{align}
g_{\mu\nu} &= \eta_{\mu\nu} + g^{(1)}_{\mu\nu} g_s + g^{(2)}_{\mu\nu}
g_s^2 +\ldots \notag\\
F_{\mu_1\dots\mu_{p+2}} &= F^{(1)}_{\mu_1\dots\mu_{p+2}} g_s + 
F_{\mu_1\dots\mu_{p+2}}^{(2)} g_s^2 +\ldots
\end{align}
where $\eta_{\mu\nu}$ is the flat space metric. The first-order 
perturbations $g^{(1)}_{\mu\nu}$ and $F^{(1)}_{\mu_1\dots\mu_{p+2}}$
correspond to the one-point amplitude
of the graviton and R-R gauge field on the disc. The second-order
metric perturbation $g^{(2)}_{\mu\nu}$ corresponds to a one-point
function of the graviton on the cylinder with the boundary state
$|S\rangle$ at the two ends \cite{Bertolini:2000jy}. It can also
be determined from the first-order perturbations by the supergravity
equations of motion, schematically:
\be 
\nabla^2 g^{(2)} = [\nabla g^{(1)}]^2 
+ \epsilon [F^{(1)}]^2 \ ,\label{so}
\ee
where $\epsilon$ is equal to $+1$ for type IIA or IIB
string theory and $-1$ for type II${}^*$ theory. This equation
is satisfied in the D-brane case, and therefore also in the $S^-$-brane
case. Since the $S^+$-brane is obtained by multiplying the R-R part
of the $S^-$ boundary state by $i$, it can only remain a solution
of \eqref{so} if the sign of $\epsilon$ is also changed. Since 
$g_{\mu\nu}^{(2)}$ involves two boundary states, there is no difficulty
in further multiplying the $S^+$-brane boundary state by a factor of 
$i$ so as to obtain purely real NS-NS and R-R fields.
The $S^+$-brane boundary state of \cite{Gutperle:2002ai} is 
thus a solution of type II* rather than type II string theory.
In contrast, the $S^-$-brane boundary state is a solution of type II 
string theory, but with non-physical imaginary fields. 
This leaves non-BPS S-branes as the only candidates of S-brane
solutions in type II string theory.

\subsubsection*{Conclusion}

In this note, we have studied some aspects of S-branes, in the 
formalism of first quantized string theory. We found that the 
instabilities in the tree-level effective action and in the one-loop
amplitude are intimately related with the infinite fine-tuning of
the initial conditions which is necessary in order to obtain an infinitely 
extended, flat S-brane. By quantizing the open strings in a spatial
direction along the S-brane, we found some evidence 
that general perturbations turn the space-like world-volume 
of the S-brane into a time-like one, which
can be interpreted as the world-volume of D-brane / anti-D-branes. 
Furthermore, we checked that Dirichlet S-branes do not exist in 
type II string theory, since it is not possible to construct a
boundary state such that the emitted Neveu-Schwarz and 
Ramond-Ramond closed string fields satisfy the correct reality conditions.
This leaves open the intriguing possibility that non-BPS S-branes, which do not
source any Ramond-Ramond field, may trigger an instability of the 
type II string theory. 

\section*{Acknowledgements}

We are grateful to Giuseppe D'Appollonio and Elias Kiritsis 
for collaboration at an early stage of this project, and 
to Costas Bachas and Rodolfo Russo for helpful discussions.


\begin{thebibliography}{00}

\bibitem{Gutperle:2002ai}
M.~Gutperle and A.~Strominger,
``Spacelike branes,''
JHEP {\bf 0204} (2002) 018
[arXiv:hep-th/0202210].

\bibitem{Sen:2004nf}
A.~Sen,
``Tachyon dynamics in open string theory,''
arXiv:hep-th/0410103.


\bibitem{Gaiotto:2003rm}
  D.~Gaiotto, N.~Itzhaki and L.~Rastelli,
  ``Closed strings as imaginary D-branes,''
  Nucl.\ Phys.\ B {\bf 688} (2004) 70
  [arXiv:hep-th/0304192].

\bibitem{Ohta:2004wk}
N.~Ohta,
``Accelerating cosmologies and inflation from M / superstring theories,''
Int.\ J.\ Mod.\ Phys.\ A {\bf 20} (2005) 1
[arXiv:hep-th/0411230].

\bibitem{Hull:1998vg}
  C.~M.~Hull,
  ``Timelike T-duality, de Sitter space, large N gauge theories and
  topological field theory,''
  JHEP {\bf 9807} (1998) 021
  [arXiv:hep-th/9806146].


\bibitem{Nekrasov:2002kf}
N.~A.~Nekrasov,
``Milne universe, tachyons, and quantum group,''
Surveys High Energ.\ Phys.\  {\bf 17} (2002) 115
[arXiv:hep-th/0203112].


\bibitem{Nitti:2005ym}
  F.~Nitti, M.~Porrati and J.~W.~Rombouts,
  ``Naturalness in cosmological initial conditions,''
  arXiv:hep-th/0503247.

\bibitem{Astefanesei:2005eq}
D.~Astefanesei and G.~C.~Jones,
``S-branes and (anti-)bubbles in (A)dS space,''
arXiv:hep-th/0502162.

\bibitem{Skenderis:2003da}
  K.~Skenderis and M.~Taylor,
  ``Properties of branes in curved spacetimes,''
  JHEP {\bf 0402} (2004) 030
  [arXiv:hep-th/0311079].

\bibitem{Okuda:2002yd}
T.~Okuda and S.~Sugimoto,
``Coupling of rolling tachyon to closed strings,''
Nucl.\ Phys.\ B {\bf 647} (2002) 101
[arXiv:hep-th/0208196].

\bibitem{Hashimoto:2004yi}
  K.~Hashimoto, P.~M.~Ho and J.~E.~Wang,
 ``Birth of closed strings and death of open strings during tachyon
  condensation,''
  Mod.\ Phys.\ Lett.\ A {\bf 20} (2005) 79
  [arXiv:hep-th/0411012].
K.~Hashimoto, P.~M.~Ho and J.~E.~Wang,
``S-brane actions,''
Phys.\ Rev.\ Lett.\  {\bf 90} (2003) 141601
[arXiv:hep-th/0211090].

\bibitem{Bachas:1992bh}
  C.~Bachas and M.~Porrati,
``Pair creation of open strings in an electric field,''
  Phys.\ Lett.\ B {\bf 296} (1992) 77
  [arXiv:hep-th/9209032].

\bibitem{Bardakci:2001ck}
K.~Bardakci and A.~Konechny,
``Tachyon condensation in boundary string field theory at one loop,''
arXiv:hep-th/0105098.

\bibitem{Craps:2001jp}
B.~Craps, P.~Kraus and F.~Larsen,
``Loop corrected tachyon condensation,''
JHEP {\bf 0106}, 062 (2001)
[arXiv:hep-th/0105227].

\bibitem{Polchinski:1995mt}
J.~Polchinski,
``Dirichlet-Branes and Ramond-Ramond Charges,''
Phys.\ Rev.\ Lett.\  {\bf 75} (1995) 4724
[arXiv:hep-th/9510017].



\bibitem{Marcus:1988vs}
N.~Marcus,
``Unitarity And Regularized Divergences In String Amplitudes,''
Phys.\ Lett.\ B {\bf 219} (1989) 265.

\bibitem{Dijkgraaf:2000fq}
  R.~Dijkgraaf, J.~M.~Maldacena, G.~W.~Moore and E.~Verlinde,
  ``A black hole farey tail,''
  arXiv:hep-th/0005003.

\bibitem{DiVecchia:1999rh}
P.~Di Vecchia and A.~Liccardo,
``D branes in string theory. I,''
NATO Adv.\ Study Inst.\ Ser.\ C.\ Math.\ Phys.\ Sci.\  {\bf 556}, 1 (2000)
[arXiv:hep-th/9912161].

\bibitem{Gaberdiel:2000jr}
M.~R.~Gaberdiel,
``Lectures on non-BPS Dirichlet branes,''
Class.\ Quant.\ Grav.\  {\bf 17} (2000) 3483
[arXiv:hep-th/0005029].

\bibitem{Bertolini:2000jy}
  M.~Bertolini, P.~Di Vecchia, M.~Frau, A.~Lerda, R.~Marotta and R.~Russo,
  ``Is a classical description of stable non-BPS D-branes possible?,''
  Nucl.\ Phys.\ B {\bf 590}, 471 (2000)
  [arXiv:hep-th/0007097].

\bibitem{DiVecchia:1997pr}
P.~Di Vecchia, M.~Frau, I.~Pesando, S.~Sciuto, A.~Lerda and R.~Russo,
``Classical p-branes from boundary state,''
Nucl.\ Phys.\ B {\bf 507} (1997) 259
[arXiv:hep-th/9707068].







    
\end{thebibliography}
\end{document}